\documentclass[12pt]{article}

\newcommand{\be}{\begin{equation}}
\newcommand{\ee}{\end{equation}}
\newcommand{\bea}{\begin{eqnarray}}
\newcommand{\eea}{\end{eqnarray}}
\newcommand{\beaa}{\begin{eqnarray*}}
\newcommand{\eeaa}{\end{eqnarray*}}

\newcommand{\g}{{\bf g}}
\newcommand{\h}{{\bf h}}
\newcommand{\kk}{{\bf k}}

\advance\textheight by \topskip
\newdimen\tableauside\tableauside=1.0ex
\newdimen\tableaurule\tableaurule=0.4pt
\newdimen\tableaustep
\def\phantomhrule#1{\hbox{\vbox to0pt{\hrule height\tableaurule width#1\vss}}}
\def\phantomvrule#1{\vbox{\hbox to0pt{\vrule width\tableaurule height#1\hss}}}
\def\sqr{\vbox{%
  \phantomhrule\tableaustep
  \hbox{\phantomvrule\tableaustep\kern\tableaustep\phantomvrule\tableaustep}%
  \hbox{\vbox{\phantomhrule\tableauside}\kern-\tableaurule}}}
\def\squares#1{\hbox{\count0=#1\noindent\loop\sqr
  \advance\count0 by-1 \ifnum\count0>0\repeat}}
\def\tableau#1{\vcenter{\offinterlineskip
  \tableaustep=\tableauside\advance\tableaustep by-\tableaurule
  \kern\normallineskip\hbox
    {\kern\normallineskip\vbox
      {\gettableau#1 0 }%
     \kern\normallineskip\kern\tableaurule}%
  \kern\normallineskip\kern\tableaurule}}
\def\gettableau#1 {\ifnum#1=0\let\next=\null\else
  \squares{#1}\let\next=\gettableau\fi\next}

\usepackage{amssymb}
\usepackage{pstricks}
\newcommand{\psl}{\psline[linewidth=0.01]}
\newcommand{\pslb}{\psline[linewidth=0.01,linestyle=dashed]}
\newcommand{\one}{\begin{pspicture}(0,.1)(.5,.6)
\psline(0,.25)(.5,.25)
\end{pspicture} }
\newcommand{\ons}{\begin{pspicture}(0,.1)(.5,.6)
\psline(0,.25)(.5,.25)
\pscircle[fillstyle=solid](.25,.25){.08}
\end{pspicture} }
\newcommand{\tws}{\begin{pspicture}(0,.1)(.5,.6)
\psline(0,0)(.5,0)
\psline(0,.5)(.5,.5)
\end{pspicture} }
\newcommand{\twc}{\begin{pspicture}(0,.1)(.5,.6)
\psline(0,0)(.5,.5)
\psline(0,.5)(.5,0)
\end{pspicture} }
\newcommand{\twssu}{\begin{pspicture}(0,.1)(.5,.6)
\psline(0,0)(.5,0)
\psline(0,.5)(.5,.5)
\pscircle[fillstyle=solid](.25,.5){.08}
\end{pspicture} }
\newcommand{\twssd}{\begin{pspicture}(0,.1)(.5,.6)
\psline(0,0)(.5,0)
\psline(0,.5)(.5,.5)
\pscircle[fillstyle=solid](.25,0){.08}
\end{pspicture} }
\newcommand{\twssb}{\begin{pspicture}(0,.1)(.5,.6)
\psline(0,0)(.5,0)
\psline(0,.5)(.5,.5)
\pscircle[fillstyle=solid](.25,0){.08}
\pscircle[fillstyle=solid](.25,.5){.08}
\end{pspicture} }
\newcommand{\ths}{\begin{pspicture}(0,.1)(.5,.6)
\psline(0,0)(.5,0)
\psline(0,.25)(.5,.25)
\psline(0,.5)(.5,.5)
\end{pspicture} }
\newcommand{\thssu}{\begin{pspicture}(0,.1)(.5,.6)
\psline(0,0)(.5,0)
\psline(0,.25)(.5,.25)
\psline(0,.5)(.5,.5)
\pscircle[fillstyle=solid](.25,.5){.08}
\end{pspicture} }
\newcommand{\thssm}{\begin{pspicture}(0,.1)(.5,.6)
\psline(0,0)(.5,0)
\psline(0,.25)(.5,.25)
\psline(0,.5)(.5,.5)
\pscircle[fillstyle=solid](.25,.25){.08}
\end{pspicture} }
\newcommand{\thssd}{\begin{pspicture}(0,.1)(.5,.6)
\psline(0,0)(.5,0)
\psline(0,.25)(.5,.25)
\psline(0,.5)(.5,.5)
\pscircle[fillstyle=solid](.25,0){.08}
\end{pspicture} }
\newcommand{\thcu}{\begin{pspicture}(0,.1)(.5,.6)
\psline(0,0)(.5,0)
\psline(0,.25)(.5,.5)
\psline(0,.5)(.5,.25)
\end{pspicture} }
\newcommand{\thcd}{\begin{pspicture}(0,.1)(.5,.6)
\psline(0,0)(.5,.25)
\psline(0,.25)(.5,0)
\psline(0,.5)(.5,.5)
\end{pspicture} }
\newcommand{\thce}{\begin{pspicture}(0,.1)(.5,.6)
\psline(0,0)(.5,.5)
\psline(0,.25)(.5,.25)
\psline(0,.5)(.5,0)
\end{pspicture} }

\makeatletter

\@addtoreset{equation}{section}

\def\section{\@startsection {section}{1}{\z@}{-3.5ex plus -1ex minus
 -.2ex}{2.3ex plus .2ex}{\large\bf\centering}}
\def\subsection{\@startsection{subsection}{2}{\z@}{-3.25ex plus%
 -1ex minus -.2ex}{1.5ex plus .2ex}{\bf}}
\def\subsubsection{\@startsection{subsubsection}{3}{\z@}{-3.25ex plus%
 -1ex minus -.2ex}{1.5ex plus .2ex}{\sl}}
\makeatother

\begin{document}

\baselineskip 17pt
\parindent 10pt
\parskip 9pt


\begin{flushright}
hep-th/0109115\\ September 2001\\[3mm]
\end{flushright}
\vspace{1cm}
\begin{center}
{\Large {\bf  Boundary remnant of Yangian symmetry and the
structure of rational reflection matrices}}\\ \vspace{1cm} {\large
G. W. Delius, N. J. MacKay, B. J. Short}
\\
\vspace{3mm} {\em Department of Mathematics,\\ University of York,
\\York YO10 5DD, U.K.\footnote{emails: {\tt gwd2, nm15, bjs108 @york.ac.uk} }}
\end{center}

\begin{abstract}
\noindent For the classical principal chiral model with boundary,
we give the subset of the Yangian charges which remains conserved
under certain integrable boundary conditions, and extract them
from the monodromy matrix. Quantized versions of these charges are
used to deduce the structure of rational solutions of the
reflection equation, analogous to the `tensor product graph' for
solutions of the Yang-Baxter equation. We give a variety of such
solutions, including some for reflection from non-trivial boundary
states, for the $SU(N)$ case, and confirm these by constructing
them by fusion from the basic solutions.
\end{abstract}


\vspace{0.5cm}
\section{The principal chiral model with boundary}

\subsection{Classical boundary conditions and conserved charges}

In a recent paper \cite{macka01} (to which the reader is referred
for more detail and references), two of us explored the classical
integrability of the principal chiral model (PCM) with boundary,
and the corresponding quantum boundary $S$-matrices. The model is
defined by the action \be\label{pcmlagr} {\cal L} = {1\over 2}{\rm
Tr}\left(
\partial_\mu g^{-1}
 \partial^\mu g\right) \, ,
\ee where the field $g(x^\mu)$ takes values in a compact Lie group
$G$, and is defined in 1+1D Minkowski spacetime with
$-\infty<x\leq 0$.

We found two classes of classical boundary condition which
preserve the conservation and involution of the charges necessary
for integrability. Here we discuss only the `chiral' condition,
\be\label{chiral} g(0)\in k_L H k_R^{-1}\,, \ee where $k_{L,R}$
are arbitrary group elements and $H$ is a maximal Lie subgroup of
$G$ such that $G/H$ is a symmetric space. For simplicity we also
set $k_L=k_R=e$, the identity element. The global $G_L\times G_R$
symmetry of the original bulk model, given by $g\mapsto UgV^{-1}$
and generated by the conserved currents \be\label{lrcurr} j_\mu^L=
\partial_{\mu} g \,g^{-1} , \qquad j_\mu^R = -
g^{-1}\partial_{\mu} g, \ee is thus broken to $H\times H$. For the
currents ($L$ or $R$), and writing the Lie algebras of $G$ and $H$
as $\g$ and $\h$ respectively, we have $j_0(0)\in \h$, while the
boundary equation-of-motion then requires $j_1(0)\in \kk$, where
$\g=\h\oplus\kk$. Alternatively, since $H$ is the subgroup of $G$
fixed under an involution $\sigma$ of $\g$, we can write
$j_0=\sigma(j_0)$, and $j_1=-\sigma(j_1)$, at $x=0$.

In the bulk model, the $G\times G$ symmetry sits inside a larger
$Y(\g)\times Y(\g)$ symmetry, where $Y(\g)$ is the Yangian
algebra. This is generated by charges (where we use the
conventions of \cite{evans99})\bea \label{Q0} Q^{(0)a} & = & \int
j_0^a \,dx\\ \label{Q1} Q^{(1)a} & =  & \int j_{1}^{a} {dx} -
{1\over 2}f^a_{\;\;bc}\int j_{0}^{b}(x) \int^{x} j_{0}^{c}(y)
\,{dy} \,{dx}\,\eea using $j^L$ and $j^R$ respectively, decomposed
into $j=j^a t_a$ where the $t^a$ are generators of $\g$ with
$[t_a,t_b]=f_{ab}^{\;\;\;c} t_c$. The integrals are over all
space, $(-\infty,\infty)$ for the bulk model. But on the half-line
$(-\infty,0]$, these charges are no longer generally conserved.
However, there are two important sets of charges which do remain
conserved. Writing $\h$-indices as $i,j,k,..$ and $\kk$-indices as
$p,q,r,...$, and noting that the only non-zero structure constants
are $f^i_{\;\;jk}$ and $f^i_{\;\;pq}$ (and cycles thereof), these
are \bea\label{Q0b} &&Q^{(0)i}\\{\rm and} && \widetilde{Q}^{(1)p}
\equiv Q^{(1)p} + {1\over 2} f^p_{\;\;qi} Q^{(0)i} Q^{(0)q}\,.
\label{Q1b} \eea The first set generates $H$ and was noted in
\cite{macka01}. To check their conservation, we note that \beaa
&&{d\over dt}\, Q^{(0)i} = j_1^i(0)=0
\\[0.1in]{\rm and}&& {d\over dt}\, \widetilde{Q}^{(1)p} = {d\over dt} Q^{(1)p} +
{1\over 2} f^p_{\;\;qi} Q^{(0)i} j_1^q(0) =0 \\[0.1in] {\rm
since}&& {d\over dt} \,Q^{(1)p} ={1\over 2} f^p_{\;\;iq} Q^{(0)i}
j_1^q(0)\,.\eeaa It is difficult to prove rigorous quantum results
for the PCM, but we shall assume that these charges remain
conserved after quantization, in the form \bea\label{Q0q}
&&Q^{(0)i}\\{\rm and} && \widetilde{Q}^{(1)p} \equiv Q^{(1)p} +
{1\over 4} [C_2^\h, Q^{(0)p}] \label{Q1q}\,. \eea Here we have set
$\hbar=1$ for convenience, and $C_2^\h\equiv \gamma_{ij}Q^{(0)i}
Q^{(0)j}$ is the quadratic Casimir operator of $\g$ restricted to
$\h$, with $\gamma_{ij}=f_{ia}^{\;\;\;b}f_{jb}^{\;\;\;a}$.

\subsection{Conserved charges from the monodromy matrix}

It is straightforward to construct these charges as coefficients
of a spectral parameter in a monodromy matrix. The conservation
and curvature-freedom of the bulk PCM currents can be expressed
through a Lax pair, $[\partial_0-L_0,\partial_1-L_1]=0$ where $$
L_1 = {1\over 1-u^2} \left(j_1-u j_0\right) \,,\qquad L_0 =
{1\over 1-u^2} \left(j_0-u j_1\right)\,.$$ The Yangian charges
then appear in the monodromy matrix $$ T^\infty_{-\infty}(u)
\equiv{\bf P}\exp\left({\int_{-\infty}^\infty L_1\,dx}\right) =
\exp\left( {1\over u} Q^{(0)a} t_a - {1\over u^2} Q^{(1)a}t_a +
\ldots \right)\,.$$ We can regard our model on the half-line, with
boundary condition $j_0=\sigma(j_0)$ and $j_1=-\sigma(j_1)$ at
$x=0$, as a restriction of the bulk model with
$j_0(x)=\sigma(j_0(-x))$ and $j_1(x)=-\sigma(j_1(-x))$. We then
have $$ T^\infty_{-\infty}(u) = \sigma\left(
(T^0_{-\infty})^{-1}(-u)\right) T^0_{-\infty}(u)\,,$$ which is
conserved because $L_0(u)=\sigma(L_0(-u))$. Expanding this gives
\beaa&& \exp\left({1\over u} Q^{(0)a} \sigma(t_a) +{1\over u^2}
Q^{(1)a}\sigma(t_a) + \ldots \right) \exp\left( {1\over u}
Q^{(0)a} t_a - {1\over u^2} Q^{(1)a}t_a + \ldots\right)\\ & = & 1+
{1\over u} Q^{(0)a}(t_a+ \sigma(t_a)) -{1\over u^2}
\left\{Q^{(1)a}(t_a-\sigma(t_a)) -Q^{(0)b} Q^{(0)c}
\left(\sigma(t_b) t_c + {1\over 2} \sigma(t_b)\sigma(t_c) +
{1\over 2} t_b t_c\right) \right\} + \ldots
\\
 & = & 1+ {2\over u} Q^{(0)i}t_i + {1\over 2} \left({2\over u}
 Q^{(0)i}t_i \right)^2
-{2\over u^2} \left(Q^{(1)p}t_p -{1\over 4} Q^{(0)b} Q^{(0)c}
[\sigma(t_b),t_c] \right) + \ldots
\\ & = & \exp\left({2\over u}Q^{(0)i}t_i -{2\over u^2} \widetilde{Q}^{(1)p}t_p
+ \ldots\right)\,, \eeaa as required.

\newpage
\section{Rational reflection matrices}

\subsection{Reflection from the boundary ground state}

Recall the structure of the rational solutions of the bulk
Yang-Baxter equation (YBE), the `$R$-matrices'. The $R$-matrix
acting on $U\otimes V$, where $U$ and $V$ are irreducible
representations of $Y(\g)$, decomposes into the sum of projectors
onto the $\g$-irreducible component representations of $U\otimes
V$. The coefficients of these projectors can then, in simple cases
-- where $U$ and $V$ are $\g$-irreducible and $U\otimes V$ has no
multiplicities --, be deduced from a `tensor product graph'
\cite{macka91} which describes how the rest of $Y(\g)$ relates
these components.

We proceed similarly now for the boundary Yang-Baxter or
`reflection' equation on $U\otimes V$, \bea
\check{R}_{VU}(\theta-\phi)( I \otimes K_U(\theta))
\check{R}_{UV}(\theta+\phi)( I \otimes K_V(\phi))=
\phantom{MMMMMMMM} \label{BYBE}
\\ \phantom{MMMMMMMM} (I \otimes
K_V(\phi))\check{R}_{VU}(\theta+\phi) (I \otimes K_U(\theta))
\check{R}_{UV}(\theta-\phi)\,,\nonumber \eea where
$\check{R}_{UV}:U\otimes V \rightarrow V \otimes U$ is a solution
of the bulk YBE, $K_U:U\rightarrow U$ is the reflection matrix and
the variables $\theta$ and $\phi$ are the rapidities of the
particles incident on the boundary. Implicit in $K$'s acting only
on the bulk multiplet $U$ or $V$ is that the boundary has no
structure of its own -- {\em i.e.\ }it is in a singlet state.
Another possibility, also discussed below, is that
$K_U:U\rightarrow \bar{U}$, in which case the reflection equation
becomes \bea \check{R}_{\bar{V}\bar{U}}(\theta-\phi)( I \otimes
K_U(\theta)) \check{R}_{U\bar{V}}(\theta+\phi)( I \otimes
K_V(\phi))= \phantom{MMMMMMMM} \label{conj}
\\ \phantom{MMMMMMMM} (I \otimes
K_V(\phi))\check{R}_{V\bar{U}}(\theta+\phi) (I \otimes
K_U(\theta)) \check{R}_{UV}(\theta-\phi)\,.\nonumber \eea In this
paper, in contrast to \cite{macka01}, we describe only the matrix
structure of (the individual $L$ and $R$ factors of) $K$, and do
not concern ourselves with the scalar prefactors and
crossing/unitarity conditions necessary construct a valid boundary
$S$-matrix \cite{ghosh94}.

Let us specialize, as in the bulk case, to $U$ and $V$ which are
$\g$-irreducible. Conservation of the $Q^{(0)i}$ requires that $$
K_U(\theta) Q^{(0)i} = Q^{(0)i} K_U(\theta) $$ (in which by
$Q^{(0)i}$ we mean its appropriate representation) and thus that
$K_U(\theta)$ act as the identity on $\h$-irreducible components
of $U$. So we have $$K_U(\theta) = \sum_{W_\h \subset U_\g}
\tau_W(\theta) P_W\,, $$ where the sum is over
$\h$-representations $W$ into which the $\g$-representation $U$
branches, and $P_W$ is the projector onto $W$. Thus the
conjugating case, $K:U\rightarrow \bar{U}$, is only admissible
when $U$ and $\bar{U}$ branch to the same $\h$-representations.

To deduce relations among the $\tau_W$ we use conservation of the
$\widetilde{Q}^{(1)p}$, recalling\footnote{Note that our
conventions differ from that of \cite{berna91}: the relative sign
of $\theta$ is due to our use of the opposite coproduct.} that, on
a $\g$-irreducible multiplet of rapidity $\theta$, the action of
$Q^{(1)}$ is given by the evaluation representation, $$ Q^{(1)a} =
\theta { c_A \over 2i\pi} Q^{(0)a} \,,$$ where $c_A$ is the value
of the quadratic Casimir of $\g$ in the adjoint representation. So
$$ K_U(\theta) \left( \theta { c_A \over 2i \pi} Q^{(0)p} +
{1\over 4}[C_2^\h, Q^{(0)p}] \right) = \left( -\theta { c_A \over
2 i\pi}Q^{(0)p} + {1\over 4}[C_2^\h, Q^{(0)p}] \right) K_U(\theta)
\,,$$ and for $W_1,W_2\subset U_\g$ such that the reduced matrix
element $\langle W_1 \vert\vert Q^{(0)p}\vert\vert W_2 \rangle\neq
0$ we have \be\label{BG} {\tau_{W_2}(\theta) \over
\tau_{W_1}(\theta)} = \left[\Delta_{12}\right]\,, \qquad{\rm
where}\footnote{This definition differs slightly from that used
in~\cite{macka01}.} \;\left[ A \right] \equiv {\frac{i\pi
A}{2c_A}+\theta \over \frac{i\pi A}{2c_A} -\theta}\ee and
$\Delta_{12} = C_2^\h(W_1)-C_2^\h(W_2)$.

To find the $W_1,W_2$ for which $\langle W_1 \vert\vert
Q^{(0)p}\vert\vert W_2 \rangle\neq 0$ we note that the $Q^{(0)p}$
(that is, the generators of $\kk$) form a representation $Z$ of
$\h$. The Wigner-Eckart theorem then requires $W_1\subset Z\otimes
W_2$. This is strictly a necessary rather than a sufficient
condition, but we shall assume its sufficiency in what
follows.\footnote{This distinction was made in the bulk case as
that between the `tensor product graph' and the `extended tensor
product graph' \cite{macka91}.}

We can therefore describe the structure of $K_U(\theta)$ by using
a `branching graph', in which the $W_i$ are the nodes, linked by
an edge, directed from $W_i$ to $W_j$ and labelled by positive
$\Delta_{ij}$, when $W_i\subset Z\otimes W_j$.

\subsection{ Example : $G=SU(N)$}

We shall examine how this works in the case of $G=SU(N)$, for
which $c_A=2N^2$. One can go through the list of all $G/H$
similarly. Case 1 below corresponds to the standard,
non-conjugating reflection equation (\ref{BYBE}), and cases 2 and
3 to the conjugating equation (\ref{conj}).

\noindent 1. $H=S(U(M)\times U(N-M))$

We denote representations of $SU(M)\times SU(N-M)$ by $(X,Y)$
where $X$ is an $SU(M)$- and $Y$ an $SU(N-M)$-representation, here
written as a Young tableau. The singlet representation is written
$1$.
 Note that $Z=
(\bar{\tableau{1}}\,,\tableau{1})\oplus(\tableau{1}\,,\bar{\tableau{1}})$
(where $\bar{\tableau{1}}$ denotes the conjugate of
$\tableau{1}$), so that the branching graphs are\beaa
U_\g=\tableau{1}:& \hspace{0.6in}&(1,\tableau{1}\,)
\stackrel{N-2M}{\longrightarrow} (\tableau{1}\,,1) \\[0.1in]
U_\g=\tableau{1 1}:&& \left(1,\tableau{1 1}\,\right)
\stackrel{N-2M-2}{\longrightarrow} (\tableau{1}\,,\tableau{1}\,)
\stackrel{N-2M+2}{\longrightarrow} \left(\tableau{1
1}\,,1\right)\\[0.1in] U_\g=\tableau{1 1 1}:&& \left(1,\tableau{1
1 1}\,\right) \stackrel{N-2M-4}{\longrightarrow}
\left(\tableau{1}\,,\tableau{1 1}\,\right)
\stackrel{N-2M}{\longrightarrow} \left(\tableau{1 1}
\,,\tableau{1}\,\right)\stackrel{N-2M+4}{\longrightarrow}
\left(\tableau{1 1 1}\,,1\right)\eeaa  and so on (in agreement
with the restriction to $SU(N)$ of the results of \cite{macka95}).

The general result, for $U$ the $r$th-rank antisymmetric tensor
($r\leq[N/2]$), and with $(p,q)$ denoting the $p$th rank $SU(M)$
and $q$th rank $SU(N-M)$ antisymmetric tensor, is easily read-off
from the graph
 $$ \hspace{-0.3in}(0,r)
\stackrel{N-2M-2(r-1)}{\longrightarrow} (1,r-1) \ldots
\stackrel{N-2M-2(r-1)+4(p-1)}{\longrightarrow}(p,r-p)\stackrel{
N-2M-2(r-1)+4p}{\longrightarrow}\ldots(r-1,1)\stackrel{N-2M+2(r-1)}{
\longrightarrow}(r,0)\,,$$ and is
 \be K_U(\theta) = \sum_{p=0}^r \prod_{q=0}^p
 [N-2M-2(r-1)+4(q-1)]P_{(p,r-p)}\,.\label{SUgen}
 \ee

\noindent 2. $H=Sp(N)\;$ ($N$ even)

$U_\g=\tableau{1}$, the vector representation, branches to the
single $ \tableau{1} $ of $Sp(N)$, and the reflection matrix is
therefore constant. Here\footnote{We continue to use Young tableau
notation for representations of $SO(N)$ and $S\!p(N)$, but it
should be understood that traces have been removed from symmetric
tableaux for $SO(N)$, and symplectic traces from antisymmetric
tableaux for $S\!p(N)$.} $Z=\tableau{1 1}\,$, and for higher
representations we have \beaa U_\g=\tableau{1 1}: &
\hspace{0.6in}& \tableau{1 1} \stackrel{N}{\longrightarrow}
1\\[0.1in] U_\g=\tableau{1 1 1}:& &\tableau{1 1 1}
\stackrel{N-2}{\longrightarrow} \tableau{1}
\\[0.1in]
U_\g=\tableau{1 1 1 1}:&&\tableau{1 1 1 1}
\stackrel{N-4}{\longrightarrow} \tableau{1 1}
\stackrel{N}{\longrightarrow} 1 \eeaa and in general for the $r$th
rank fundamental tensor of $SU(N)$, and denoting by $(p)$ the
$p$th rank antisymmetric tensor of $Sp(N)$, $$ (r)
\stackrel{}{\longrightarrow}(r-2)\ldots
\stackrel{N-2(r-2p)}{\longrightarrow} (r-2p)
\stackrel{N+4-2(r-2p)}{\longrightarrow}\ldots
\left\{\begin{array}{llll} (2)\stackrel{N}{\longrightarrow} & (0)
& \qquad$r$\;{\rm even}\\(3) \stackrel{N-2}{\longrightarrow} & (1)
& \qquad$r$\;{\rm odd}\end{array}\right.$$

\pagebreak \noindent 3. $H=SO(N)$

Here $Z=\tableau{2}$ and each reflection matrix is constant, since
the $r$th rank fundamental antisymmetric tensor representation of
$SU(N)$ branches to the same, irreducible representation of
$SO(N)$.

\subsection{Reflection from a boundary bound state}

When a reflection matrix is used to construct a boundary
$S$-matrix, it may have a pole at one of the labels of the
branching graph. We then expect a  multiplet of boundary states to
exist which transforms in an $\h$-representation corresponding to
a subgraph. The results of the previous section can be extended to
accommodate such non-trivial boundary states.

Suppose we wish to calculate the reflection matrix
$K_{U}^{[V]}(\theta)$ of bulk multiplet $U_\g$ off boundary
multiplet $V_\h$.   The $Q^{(0)i}$ have trivial coproduct, and
their conservation enforces the decomposition of
$K_{U}^{[V]}(\theta)$ into projectors $P_W$ onto the
$\h$-irreducible components $W_\h$ of $U_\g\otimes V_\h$,
$$K_U^{[V]}(\theta) = \sum_{W_\h \subset U_\g\otimes V_\h}
\tau_W(\theta) P_W\,. $$

To deduce the $\tau_W$, we need to compute the action of
 $\widetilde{Q}^{(1)p}$ on the spaces $W$. Using the Yangian
coproduct $$ \Delta(Q^{(1)a}) = Q^{(1)a}\otimes 1 + 1 \otimes
Q^{(1)a} + {1\over 2} f^a_{\;\;bc} Q^{(0)b}\otimes Q^{(0)c}\,,$$
we have \beaa \Delta(\widetilde{Q}^{(1)p}) & = &
\Delta\left(Q^{(1)p}) +{1\over 4} [C_2^\h,Q^{(0)p}]\right)\\ & = &
Q^{(1)p}\otimes 1 + 1\otimes Q^{(1)p} + {1\over
4}[C_2^\h,Q^{(0)p}]\otimes 1 + 1 \otimes  {1\over
4}[C_2^\h,Q^{(0)p}]\\&& + {1\over 2} f^p_{\;\;iq} Q^{(0)i}\otimes
Q^{(0)q} +{1\over 2} f^p_{\;\;qi} Q^{(0)q}\otimes Q^{(0)i}+{1\over
2}[\gamma_{ij}Q^{(0)i}\otimes Q^{(0)j}, Q^{(0)p}\otimes 1 + 1
\otimes Q^{(0)p}] \\ & = & \widetilde{Q}^{(1)p} \otimes 1 + 1
\otimes \widetilde{Q}^{(1)p} + [\gamma_{ij}Q^{(0)i}\otimes
Q^{(0)j}, Q^{(0)p}\otimes 1]
\\ & = &\widetilde{Q}^{(1)p} \otimes 1 + 1 \otimes \widetilde{Q}^{(1)p} +
{1\over 2}[\Delta(C_2^\h)-C_2^\h\otimes 1 - 1 \otimes C_2^\h,
Q^{(0)p}\otimes 1]\,. \eeaa Thus when $U$ is $\g$-irreducible and
$V$ is $\h$-irreducible, we have \beaa K_U^{[V]}(\theta) \left(
\theta { c_A \over 2i \pi}Q^{(0)p}\otimes 1 + {1\over
4}[C_2^\h\otimes 1, Q^{(0)p}\otimes 1] + {1\over 2}
[\Delta(C_2^\h)-C_2^\h\otimes 1 - 1 \otimes C_2^\h,
Q^{(0)p}\otimes 1] \right)\\ \;\; = \left(- \theta { c_A \over 2i
\pi}Q^{(0)p}\otimes 1 + {1\over 4}[C_2^\h \otimes 1,
Q^{(0)p}\otimes 1]+ {1\over 2} [\Delta(C_2^\h)-C_2^\h\otimes 1 - 1
\otimes C_2^\h, Q^{(0)p}\otimes 1]\right) K_U^{[V]}(\theta)
\,.\eeaa So a graph is generated exactly as for the boundary
ground state, but now with $$ {\tau_{W_2}(\theta) \over
\tau_{W_1}(\theta)} = \left[\Delta_{12}\right]\,.$$ Here
$$\Delta_{12} = 2\left( C_2^\h(W_1)-C_2^\h(W_2)\right) -
\left(C_2^\h(\widetilde{W}_1)-C_2^\h(\widetilde{W}_2)\right)\,,$$
where $\widetilde{W}_i$ is the $\h$-component of $U_\g$ from which
$W_i$  descends. Note that this subsumes the results of section
2.1: if the boundary is in the ground state,
$W_i=\widetilde{W}_i$, and we reproduce (\ref{BG}).

\subsection{Example : $G=SU(N)$, $H=S(U(M)\times U(N-M))$}

We denote by $K_p^{[q]}$ the reflection of the bulk $p$th
antisymmetric tensor off the boundary state in the $q$th
antisymmetric tensor of $SU(M)$. Then

\noindent $K_1^{[1]}:$ \qquad $
 \{(\tableau{1}\,,1) \oplus (1,
\tableau{1}\,)\}\otimes(\tableau{1}\,,1)$ $$
\left(\tableau{2}\,,1\right) \stackrel{N-2M-4}{\longleftarrow}
(\tableau{1},\tableau{1}\,) \stackrel{N-2M+4}{\longrightarrow}
\left(\tableau{1 1}\,,1\right)$$

\noindent $K_1^{[2]}:$ \qquad $ \{(\tableau{1}\,,1) \oplus (1,
\tableau{1}\,)\}\otimes(\tableau{1 1}\,,1)$ $$ \left(\tableau{2
1}\,,1\right) \stackrel{N-2M-4}{\longleftarrow} (\tableau{1
1},\tableau{1}\,) \stackrel{N-2M+8}{\longrightarrow}
\left(\tableau{1 1 1}\,,1\right)$$

\noindent $K_2^{[1]}:$ \qquad $ \{(\tableau{1 1}\,,1) \oplus
(\tableau{1}\,, \tableau{1}\,) \oplus(1,\tableau{1 1}\,)
\}\otimes(\tableau{1}\,,1)$

$$ \begin{array}{ccc}(\tableau{1}\,,\tableau{1 1}\,) &
\stackrel{N-2M-6}{\longrightarrow}
&\left(\tableau{2}\,,\tableau{1}\,\right)
 \\[0.1in] \hspace{0.35in}\downarrow_{N-2M+2} &&
\hspace{0.35in}\downarrow_{N-2M+2} \\[0.1in] (\tableau{1
1}\,,\tableau{1}\,) & \stackrel{N-2M-6}{\longrightarrow} &
\left(\tableau{2 1}\,,1\right)  \\[0.1in]
\hspace{0.35in}\downarrow_{N-2M+6}&&\\[0.1in] \left(\tableau{1 1
1}\,,1\right) &&
\end{array}$$
and so on; the result is easily generalized.

\pagebreak
\section{Fusion of reflection matrices}

For the low rank examples that we have given in the previous
section it is possible to construct the $K$-matrices explicitly by
fusion \cite{mezin92,macka95}. In this section we explain how this
is done.

\subsection{$G=SU(N)$, $H=S(U(M)\times
U(N-M))$}

We start from the vector reflection matrix $K_1^{[0]}(\theta)$.
For $H=S(U(M)\times U(N-M))$ the matrix part of this is
$$P_1+\left[N-2M\right]P_2\qquad{\rm where} \qquad
P_1={\textstyle\frac{1}{2}}(I-E)\,,\qquad
P_2={\textstyle\frac{1}{2}}(I+E)\,.$$ Here $E$ is an $N\times N$
matrix, determined by the boundary conditions~\cite{macka01},
which satisfies $E^2=I_N$ (the $N\times N$ identity matrix), so
that the above are orthogonal projectors. (Recall \cite{macka01}
that the admissible $E$ locally parametrize $G/H$, although
choosing $k_L=k_R=1$ makes the specific choice $E=$
diag$(I_M,-I_{N-M})$.) They project onto $(1,\tableau{1}\,)$ and
$(\tableau{1}\,,1)$ respectively, so that the vector particle
reflection matrix is as given in the previous section.

We calculate the second rank reflection matrix $K_2^{[0]}$ by
fusing together two vector multiplets. The vector multiplet bulk
$S$-matrix has a simple pole at $\theta=\frac{2i\pi}{c_A}$ whose
residue is the projector onto the second-rank antisymmetric
representation of $SU(N)$. This is interpreted according to the
bootstrap principle as another particle multiplet, for which  the
reflection matrix is \bea K_2^{[0]}(\theta)= \check{R}_{11}(2i\pi/
N)\left( I \otimes K_1^{[0]}(\theta+i\pi/N)\right)
\check{R}_{11}(2\theta)\left( I \otimes K_1^{[0]}(\theta-i\pi/
N)\right)\,,\eea or, diagrammatically,\\

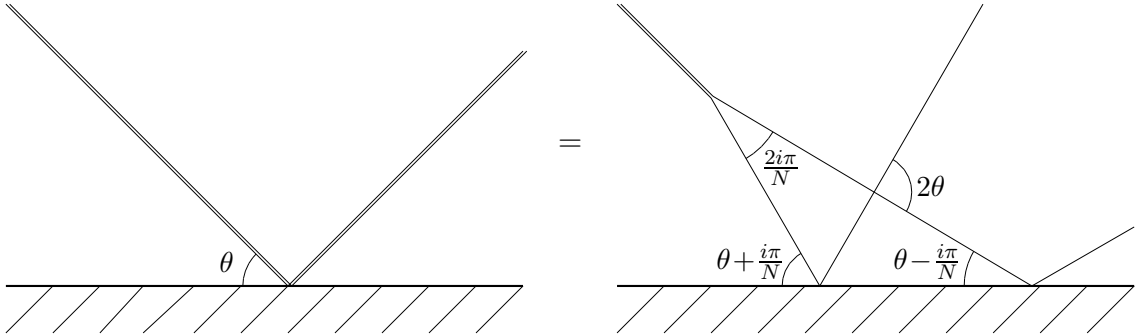
\begin{pspicture}(-.55,0)(15,4.875)
\psline(0,.625)(6.875,.625) \psline(8.125,.625)(15,.625)
\psl(0,0)(.625,.625) \psl(.625,0)(1.25,.625)
\psl(1.25,0)(1.825,.625) \psl(1.825,0)(2.5,.625)
\psl(2.5,0)(3.125,.625) \psl(3.125,0)(3.75,.625)
\psl(3.75,0)(4.375,.625) \psl(4.375,0)(5,.625)
\psl(5,0)(5.625,.625) \psl(5.625,0)(6.25,.625)
\psl(6.25,0)(6.875,.625) \psl(8.125,0)(8.75,.625)
\psl(8.75,0)(9.375,.625) \psl(9.375,0)(10,.625)
\psl(10,0)(10.625,.625) \psl(10.625,0)(11.25,.625)
\psl(11.25,0)(11.875,.625) \psl(11.875,0)(12.5,.625)
\psl(12.5,0)(13.125,.625) \psl(13.125,0)(13.75,.625)
\psl(13.75,0)(14.375,.625) \psl(14.375,0)(15,.625)
\psl(0,4.375)(3.75,.625) \psl(.05,4.375)(3.8,.625)
\psl(3.75,.625)(6.875,3.75) \psl(3.8,.625)(6.925,3.75)
\psl(3.15,.625)(3.1584,.725) \psl(3.1584,.725)(3.1843,.825)
\psl(3.1843,.825)(3.2304,.925) \psl(3.2304,.925)(3.2816,1)
\psl(3.2816,1)(3.3257,1.0493) \rput(2.938,.95){{\small $\theta$}}
\psl(8.125,4.375)(9.375,3.125) \psl(8.175,4.375)(9.4,3.15)
\psl(9.375,3.125)(10.82,.625) \psl(9.4,3.15)(13.64,.625)
\psl(10.82,.625)(12.99,4.375) \psl(13.64,.625)(15,1.41)
\psl(10.0453,2.4797)(9.9382,2.3868)
\psl(10.0453,2.4797)(10.1382,2.5868)
\psl(9.9382,2.3868)(9.8386,2.321)
\psl(10.1382,2.5868)(10.1974,2.6751) \rput(10.3,2.2){{\small
$\frac{2i\pi}{N}$}} \psl(10.32,.625)(10.3301,.725)
\psl(10.3301,.725)(10.3617,.825) \psl(10.3617,.825)(10.42,.925)
\psl(10.42,.925)(10.52,1.025) \psl(10.52,1.025)(10.5698,1.0579)
\rput(9.9,.95){{\small $\theta\!+\!\frac{i\pi}{N}$}}
\psl(12.74,.625)(12.7456,.725) \psl(12.7456,.725)(12.7625,.825)
\psl(12.7625,.825)(12.7915,.925) \psl(12.7915,.925)(12.8338,1.025)
\psl(12.8338,1.025)(12.8667,1.0855) \rput(12.25,.95){{\small
$\theta\!-\!\frac{i\pi}{N}$}} \psl(12.0427,1.874)(12.0326,1.974)
\psl(12.0427,1.874)(12.0326,1.774)
\psl(12.0326,1.974)(12.001,2.074)
\psl(12.0326,1.774)(12.001,1.674)
\psl(12.001,2.074)(11.9427,2.174)
\psl(12.001,1.674)(11.9723,1.6182)
\psl(11.9427,2.174)(11.8427,2.274)
\psl(11.8427,2.274)(11.7931,2.3068) \rput(12.3,1.95){{\small
$2\theta$}} \rput(7.5,2.5){$=$}
\end{pspicture} \\
We now introduce the diagrammatic notation $I=\one $ and $E=\ons
$, so that, for example, the permutation operator on two vectors
is written $\twc$. (The reader is referred to~\cite{macka01} for
further details.) Expressing the above diagram in this notation,
we obtain
$$K_2^{[0]}(\theta)\propto\left(\tws-\twc\right)\left(\tws+c\left(\theta+{\textstyle\frac{i\pi}
{N}}\right)\twssd\right)\left(\tws-{\textstyle\frac{N\theta}
{i\pi}}\twc\right)\left(\tws+c\left(\theta-{\textstyle\frac{i\pi}{N
}}\right)\twssd\right)\,,$$ where $c={2N\over i\pi(N-2M)}$.
Expanding this, up to an overall scalar factor we obtain
$$K_2^{[0]}\propto P_2^A\left(P_1+\left[N-2M-2\right]\left(
P_2+\left[N-2M+2\right]P_3\right)\right)$$ where
$P_2^A=\frac{1}{2}\left(\tws-\twc\right)$ is the projector onto
the second rank antisymmetric representation of $SU(N)$ and
$P_{1,2,3}$ are the orthogonal projectors
$$P_1=\frac{1}{4}\left(\tws-\twssu\right)\left(\tws-\twssd\right)\,,\quad
P_2=\frac{1}{2}\left(\tws-\twssb\right)\quad{\rm and} \quad
P_3=\frac{1}{4}\left(\tws+
\twssu\right)\left(\tws+\twssd\right)\,.$$  $P_2^A P_{1,2,3}$
project onto the irreducible representations $ \left(1,\tableau{1
1}\,\right)$, $(\tableau{1}\,,\tableau{1}\,)$ and
$\left(\tableau{1 1} \,,1\right)$ respectively, in agreement with
the previous section. Note that these explicit expressions for the
projectors make it clear how the $\h$-representations are embedded
into the parent $SU(N)$ representation $\tableau{1 1}$. A similar
but more complicated calculation gives the matrix part of
$K_3^{[0]}(\theta)$ in the same way. We also have an inductive
construction of the matrix part of $K_n^{[0]}$ which reproduces
(\ref{SUgen}).

\subsection{$G=SU(N)$, $H=S\!p(N)$ and $H=SO(N)$}

For these cases (in which the `conjugated' reflection equation
(\ref{conj}) applies) the matrix part $E$ of $K_1^{[0]}$ is
constant, and is symmetric in the $SO(N)$ and antisymmetric in the
$S\!p(N)$ case. We have again performed the fusion calculations of
$K_2^{[0]}$ (see also \cite{macka95}) and $K_3^{[0]}$, and
reproduced the results of section two.

Recall that for $H=SO(N)$ all the $K$-matrices are constant,
whereas for $H=S\!p(N)$, due to the non-trivial branching rule,
extra structure appears. In the fusion calculation, this
distinction becomes apparent when $P_2^A$ is contracted with $E$:
for the symmetric, $SO(N)$ case, such terms vanish, whereas for
the antisymmetric, $S\!p(N)$ case, they give non-trivial
contributions.

\subsection{Scattering off an excited boundary}

We return now to the Grassmannian case $H=S(U(M)\times U(N-M))$
and consider boundary scattering off excited boundary states. It
was noted in~\cite{macka01} that the vector particle boundary
scattering matrix $K_1^{[0]}$ has a simple pole at
$\theta=\frac{(N-2M)i\pi}{2N}=\theta_0$ corresponding to the
formation of a boundary bound state, transforming in
$(\tableau{1}\,,1)$.

We determine the reflection matrix $K_1^{[1]}(\theta)$ of the
vector multiplet off this state as follows. The bootstrap
principle, applied now to the {\em boundary} bound state, gives
\bea K_1^{[1]}(\theta) =(I \otimes
K_1^{[0]}(\theta_0))\check{R}_{11}(\theta+\theta_0) (I \otimes
K_1^{[0]}(\theta)) \check{R}_{11}(\theta-\theta_0)\,, \eea or,
diagrammatically,
\\
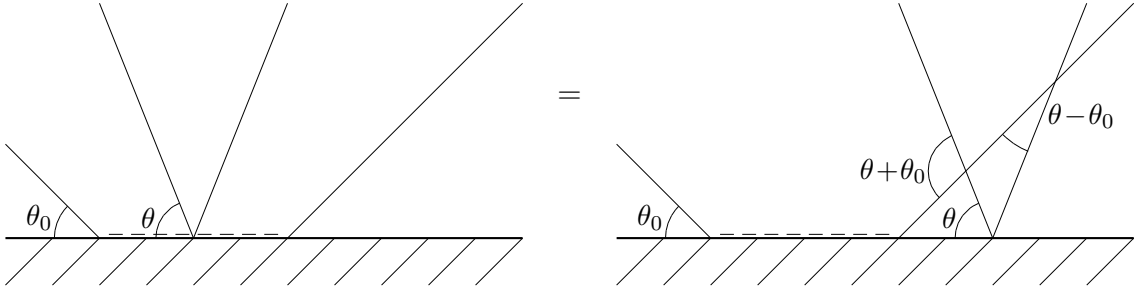
\begin{pspicture}(-.55,0)(15,4.25)
\psline(0,.625)(6.875,.625)
\psline(8.125,.625)(15,.625)
\psl(0,0)(.625,.625)
\psl(.625,0)(1.25,.625)
\psl(1.25,0)(1.875,.625)
\psl(1.875,0)(2.5,.625)
\psl(2.5,0)(3.125,.625)
\psl(3.125,0)(3.75,.625)
\psl(3.75,0)(4.375,.625)
\psl(4.375,0)(5,.625)
\psl(5,0)(5.625,.625)
\psl(5.625,0)(6.25,.625)
\psl(6.25,0)(6.875,.625)
\psl(8.125,0)(8.75,.625)
\psl(8.75,0)(9.375,.625)
\psl(9.375,0)(10,.625)
\psl(10,0)(10.625,.625)
\psl(10.625,0)(11.25,.625)
\psl(11.25,0)(11.875,.625)
\psl(11.875,0)(12.5,.625)
\psl(12.5,0)(13.125,.625)
\psl(13.125,0)(13.75,.625)
\psl(13.75,0)(14.375,.625)
\psl(14.375,0)(15,.625)
\rput(7.5,2.5){$=$}
\psl(0,1.875)(1.25,.625)
\pslb(1.375,.675)(3.625,.675)
\psl(3.75,.625)(6.875,3.75)
\psl(.65,.625)(.6584,.725)
\psl(.6584,.725)(.6843,.825)
\psl(.6843,.825)(.7304,.925)
\psl(.7304,.925)(.7816,1)
\psl(.7816,1)(.8257,1.0493)
\rput(.45,.9){{\small $\theta_0$}}
\psl(1.25,3.75)(2.5,.625)
\psl(2.5,.625)(3.75,3.75)
\psl(2,.625)(2.0101,.725)
\psl(2.0101,.725)(2.0417,.825)
\psl(2.0417,.825)(2.1,.925)
\psl(2.1,.925)(2.2,1.025)
\psl(2.2,1.025)(2.2366,1.05)
\psl(2.2366,1.05)(2.3143,1.0892)
\rput(1.9,.9){{\small $\theta$}}
\psl(8.775,.625)(8.7834,.725)
\psl(8.7834,.725)(8.8093,.825)
\psl(8.8093,.825)(8.8554,.925)
\psl(8.8554,.925)(8.9066,1)
\psl(8.9066,1)(8.9507,1.0493)
\rput(8.575,.9){{\small $\theta_0$}}
\psl(8.125,1.875)(9.375,.625)
\pslb(9.5,.675)(11.75,.675)
\psl(11.875,.625)(15,3.75)
\psl(11.875,3.75)(13.125,.625)
\psl(13.125,.625)(14.375,3.75)
\psl(12.625,.625)(12.6351,.725)
\psl(12.6351,.725)(12.6667,.825)
\psl(12.6667,.825)(12.725,.925)
\psl(12.725,.925)(12.825,1.025)
\psl(12.825,1.025)(12.8616,1.05)
\psl(12.8616,1.05)(12.9393,1.0892)
\rput(12.55,.9){{\small $\theta$}}
\psl(12.2679,1.5179)(12.278,1.6179)
\psl(12.2679,1.5179)(12.278,1.4179)
\psl(12.278,1.6179)(12.3096,1.7179)
\psl(12.278,1.4179)(12.3096,1.3179)
\psl(12.3096,1.7179)(12.3679,1.8179)
\psl(12.3096,1.3179)(12.3679,1.2179)
\psl(12.3679,1.2179)(12.4143,1.1643)
\psl(12.3679,1.8179)(12.4679,1.9179)
\psl(12.4679,1.9179)(12.5045,1.9429)
\psl(12.5045,1.9429)(12.5822,1.9921)
\rput(11.8,1.52){{\small $\theta\!+\!\theta_0$}}
\psl(13.2612,2.0012)(13.3,1.9644)
\psl(13.3,1.9644)(13.4,1.8855)
\psl(13.4,1.8855)(13.5,1.8247)
\psl(13.5,1.8247)(13.5969,1.7798)
\rput(14.3,2.25){{\small $\theta\!-\!\theta_0$}}
\end{pspicture} \\
This confirms the results of section 2.4, with explicit
expressions for the projectors as follows: for $K_1^{[1]}$ we have
\beaa P_{(\tableau{1}\,,\tableau{1}\,)} & = &
\frac{1}{4}\left(\tws+ \twssd\right)\left(\tws-\twssu\right)\\
P_{(\tableau{2}\,,1)} & = &
\frac{1}{8}\left(\tws+\twc\right)\left(\tws+\twssd\right)\left(\tws+
\twssu\right)\\ P_{\left(\tableau{1 1}\,,1\right)} & = &
\frac{1}{8}
\left(\tws-\twc\right)\left(\tws+\twssd\right)\left(\tws+\twssu\right)
\,.\nonumber \eeaa Proceeding similarly for $K_1^{[2]}$ we
obtain\beaa P_{\left(\tableau{1 1}\,,\tableau{1}\,\right)} & = &
\frac{1}{16}\left(\ths-\thcd\right)\left(\ths+\thssm\right)\left(
\ths+\thssd\right)\left(\ths-\thssu\right)
\\P_{\left(\tableau{2 1}\,,1\right)} & = & \frac{1}{48}\left(\ths-\thcd
\right)\left(2\ths+\thcd+\thce\right)\left(\ths+\thssm\right)\left(
\ths+\thssd\right)\left(\ths+\thssu\right)
\\P_{\left( \tableau{1 1 1}\,,1\right)} & = & \frac{1}{48}\left(\ths-
\thcd\right)\left(\ths-\thcd-\thce\right)\left(\ths+\thssm\right)\left(
\ths+\thssd\right)\left(\ths+\thssu\right)\,,\nonumber\eeaa and
for $K_2^{[1]}$ \beaa P_{\left(\tableau{1}\,,\tableau{1
1}\,\right)} & = & \frac{1}{16}\left(
\ths-\thcu\right)\left(\ths+\thssd\right)\left(\ths-\thssu\right)\left(
\ths-\thssm\right) \\P_{(\tableau{2}\,,\tableau{1}\,)} & = &
\frac{1}
{16}\left(\ths-\thcu\right)\left(\ths+\thcd\right)\left(\ths+\thssd
\right)\left(\ths-\thssu\right)\left(\ths+\thssm\right)
\\P_{\left( \tableau{1 1}\,,\tableau{1}\,\right)} & = &
\frac{1}{32}\left(\ths-\thcu
\right)\left(\ths-\thcd\right)\left(\ths+\thssd\right)\left(\ths-\thssu
\right)\left(\ths+\thssm\right) \\ P_{\left(\tableau{2 1}\,,1
\right)} & = &
\frac{1}{48}\left(\ths-\thcu\right)\left(2\ths+\thcd+\thce\right)
\left(\ths+\thssd\right)\left(\ths+\thssu\right)\left(\ths+\thssm\right)
\\ P_{\left(\tableau{1 1 1}\,,1\right)} & = & \frac{1}{48}\left(\ths-
\thcu\right)\left(\ths-\thcd-\thce\right)\left(\ths+\thssd\right)\left(
\ths+\thssu\right)\left(\ths+\thssm\right) \,.\nonumber\eeaa

\section{Concluding remarks}

The natural next step is to put all these results together to find
the full set of boundary $S$-matrices in the PCM, and thereby the
complete spectrum of boundary bound states and their interactions.
Fusion calculations rapidly become intractable with increasing
rank, but with these in combination with the graphical methods
developed here we hope to be able to progress towards the
completion of this programme.

There also remains much to be discovered at the classical level --
for example, about what happens when a Wess-Zumino term is added
to the PCM action, to make contact with work on D-branes in group
manifolds. From the mathematical point of view it remains to
integrate our results with those on representations of twisted
Yangians \cite{molev96}. It also remains to understand how these
results apply in Gross-Neveu models, and their relationship with
the results of \cite{sorba01}.

An immediate prospect, and work in progress, is to apply similar
ideas to the trigonometric case, where the underlying model is
affine Toda field theory. We have constructed the remnant of the
quantum affine algebra symmetry generated by non-local charges in
\cite{delius01}.

\vskip 20pt {\bf Acknowledgments}

We thank Evgueni Sklyanin for discussions, and the UK EPSRC for
BJS's studentship and GWD's advanced fellowship. NJM thanks the
Nuffield foundation for a grant.

\end{document}